**Efficient Second-Harmonic Generation from Silicon Slotted Nanocubes with Bound States in the Continuum**

*Cizhe Fang, Qiyu Yang, Qingchen Yuan, Linpeng Gu, Xuetao Gan,\* Yao Shao, Yan Liu,\* Genquan Han, and Yue Hao*

C. Fang, Q. Yang, Prof. Y. Liu, Prof. G. Han, Prof. Y. Hao
Wide Bandgap Semiconductor Technology Disciplines State Key Laboratory, School of Microelectronics, Xidian University, Xi'an, 710071, China
E-mail: xdliuyan@xidian.edu.cn

Q. Yuan, L. Gu, Prof. X. Gan
Key Laboratory of Light Field Manipulation and Information Acquisition, Ministry of Industry and Information Technology, and Shaanxi Key Laboratory of Optical Information Technology, School of Physical Science and Technology, Northwestern Polytechnical University, 710129 Xi'an, China
E-mail: xuetaogan@nwpu.edu.cn

Y. Shao
Shanghai Energy Internet Research Institute of State Grid, 251 Libing Road, Pudong New Area, Shanghai, 201210, China.

Prof. Y. Liu
The Research Center for Intelligent Chips and Devices Zhejiang Lab, Hangzhou, 311121, China

Optical materials with centrosymmetry, such as silicon and germanium, are unfortunately absent of second-order nonlinear optical responses, hindering their developments in efficient nonlinear optical devices. Here, a design with an array of slotted nanocubes is proposed to realize remarkable second harmonic generation (SHG) from the centrosymmetric silicon, which takes advantage of enlarged surface second-order nonlinearity, strengthened electric field over the surface of the air-slot, as well as the resonance enhancement by the bound states in the continuum. Compared with that from the array of silicon nanocubes without air-slots, SHG from the slotted nanocube array is improved by more than two orders of magnitude. The experimentally measured SHG efficiency of the silicon slotted nanocube array is high as $1.8\times10^{-4}$ W$^{-1}$, which is expected to be further engineered by modifying the air-slot geometries.

Our result could provide a new strategy to expand nonlinear optical effects and devices of centrosymmetric materials.

## 1. Introduction

Silicon is one of the most promising materials for developing advanced photonic[1]-[3] and optoelectronic devices.[4],[5] First, its high refractive index allows the compact footprint of silicon-based photonic integrated circuits,[6] which also supports multiple Mie resonance modes for versatile meta-devices.[7],[8] Second, beneficial from the mature complementary metal-oxide-semiconductor (CMOS) technology, silicon photonic devices have great advantages of low-cost and large-scale productions over other materials, including III-V compound semiconductors,[9],[10] lithium niobate,[11],[12] polymers, etc. Third, it is possible to combine silicon photonic devices with well-developed silicon microelectronic devices for constructing optoelectronic integrated circuits. Remarkably, silicon has significant third-order nonlinear optical responses,[13],[14] providing an extra degree of freedom to expand the functions of photonic devices, such as all-optical switch,[15] generation of entangled photons,[16],[17] phase conjugation,[18] and so on. In nonlinear optics,[19]-[22] the second-order nonlinearity ($\chi^{(2)}$) has susceptibilities about ten orders of magnitude higher than third-order nonlinear susceptibilities ($\chi^{(3)}$), which is much preferred in typical nonlinear photonic devices.[23],[24] Unfortunately, silicon is a centrosymmetric crystal, making it absent of bulk second-order optical nonlinearity.[25]-[27]

The broken centrosymmetry is allowed at silicon surface or interface, enabling a possibility to realize silicon-based second-order nonlinear optical responses. Unfortunately, the surface second-order nonlinearity only happens in few-atom thicknesses. The corresponding nonlinear responses are very weak due to the ineffective light-matter coupling.[28] The employment of silicon nanoparticles[25] or nanowires[29] with large surface-to-volume is one solution to improve the second-order nonlinearity by effectively enlarging the surface area. Another way is the utilization of optical resonators, which could confine light for a long time to effectively interact with the silicon surface, like microrings,[30] photonic crystal cavities,[26],[31] Mie resonators.[13],[14] Recently, it was revealed that SHG can be observed in high-$Q$ Si metasurfaces with broken in-plane inversion symmetry.[32] This approach opens new perspectives for SHG physics and device applications. However, the efficiencies of second-order nonlinear responses from silicon realized in these strategies still have rooms to be improved.

In this work, we demonstrate that surface second-order nonlinearity of silicon could be greatly improved by fabricating an array of slotted nanocubes, giving rise to an efficient SHG. It benefits from the simultaneous employment of surface nonlinearity and optical resonance. The air-slot in the nanocube not only enlarges the surface area with second-order nonlinearity, but also boosts the surface optical field governed by the continuous condition of the normal electric displacement. In addition, by arranging the slotted nanocubes into an array, quasi-bound states in the continuum (BIC) modes with high quality ($Q$) factor are formed, which could localize the optical field around the silicon surface for a long time for effective light-matter interaction.

## 2. Structure Design and Theoretical Analysis

**Figure 1**a schematically displays a silicon nanocube without air-slot located on a sapphire substrate. It has a thickness of $t$, and a side length of $l$. Owning to silicon's high refractive index, the nanocube could support Mie resonance modes though its dimension may be smaller than the resonance wavelength.[33] With the mode simulation based on a finite element method (see Experimental Section for details), a magnetic-dipole (MD) type Mie resonance mode of the silicon nanocube is obtained, as shown in Figure 1b. With $l$=440 nm, the resonance wavelength locates at 1687 nm, and the quality ($Q$) factor is 5.7. The electric field of the resonance mode is mostly localized inside the nanocube with a doughnut spatial distribution, which has no contribution to the SHG due to the absence of bulk $\chi^{(2)}$. While there exist mode leakages at the boundaries of the nanocube, the electric field is much weaker than the bulk counterpart, which can not yield strong surface SHG.

To solve the awkwardness, an air-slot is introduced in the middle of the nanocube, as schematically shown in Figure 1c, which has a width of *a* and a depth of *b*. The slotted nanocube could still support the Mie resonance. With *a*=110 nm, *b*=110 nm, an MD-type resonance mode is obtained at the wavelength of 1555 nm, which is blue-shifted with respect to the nanocube without air-slot. The *Q* factor is about 5.3 and the corresponding mode distribution is displayed in Figure 1d. A strong electric field emerges inside the air-slot, which is about 3 times higher than the maximum value inside the nanocube. This could be attributed to the continuous condition of the normal component of the electric displacement at the interface (i.e. $\varepsilon_0 E_{air}=\varepsilon_{Si}E_{Si}$).[34] Given silicon's high refractive index, the electric field can be obviously enhanced at the air-slot boundaries. This strengthening could be improved significantly by arranging the slotted nanocubes into an array, as discussed below. From the comparison, there are two merits for facilitating surface SHG: (*i*) the electric field interacting with the surface is significantly increased around the air-slot; (*ii*) the surface area is enlarged for providing more surface $\chi^{(2)}$.

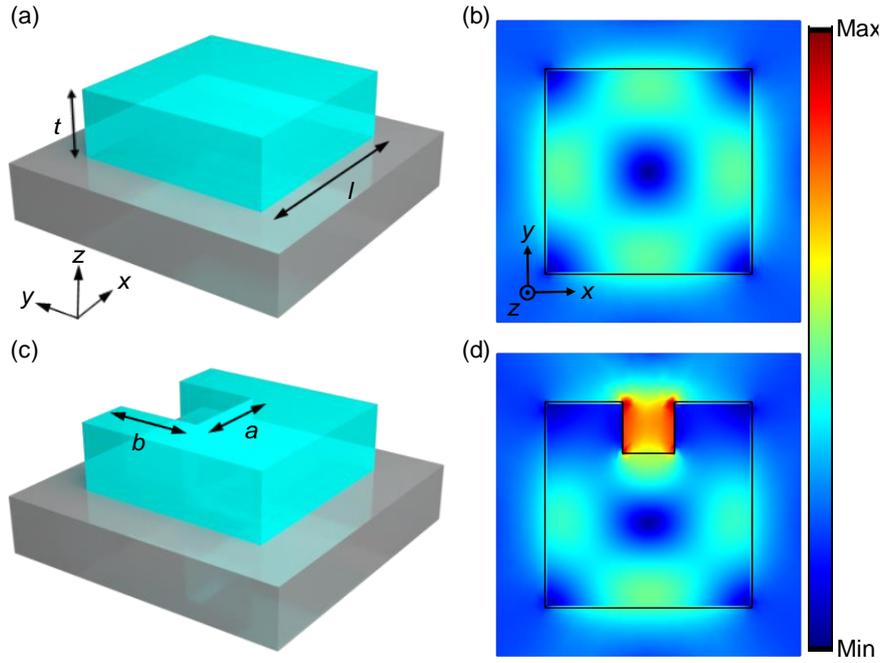

**Figure 1.** (a, b) Structure and mode distribution of a silicon nanocube without air-slot. (c, d) Structure and mode distribution of a silicon nanocube with an air-slot.

For a single slotted nanocube, the $Q$ factor of the Mie resonance mode is still small. In an optical cavity, the localized electric field is proportional to the $Q$ factor. To realize stronger SHG from the slotted nanocube, the $Q$ factor should be improved greatly. In addition, by comparing the localized electric fields in the single nanocubes with and without air-slot (shown in Figures 1b and 1d), the boosting effect on the electric field by the air-slot is not remarkable enough, which is caused by the strong mode leaking of the single nanocube. If the $Q$ factor could be improved to reduce the mode leaking, the air-slot-assisted strengthening on the electric field is expected to present. As schematically shown in the inset of **Figure 2**a, slotted nanocubes are arranged into an array to improve the $Q$ factor. In the array, the Mie resonance mode of individual nanocubes would couple with each other, which provides a possibility to decrease the far-field radiation with the destructive interference among the leaky modes. Furthermore, from the view of future developments in practical nonlinear optical devices, an array of nanocubes is more promising than a single nanocube. It is a challenge to focus an optical beam into a size matching with the single subwavelength nanocube. With the slotted nanocubes in an array, the device dimension could be designed large enough to be compatible with the beam size of the excitation laser. Finally, the large nanocube array could provide much more silicon surfaces second-order nonlinearity to enhance SHG than a single nanocube.

Figure 2a shows the calculated $Q$ factors of the resonance modes from the slotted nanocube array with different array sizes. Here, the slotted nanocube has parameters of $l$=440 nm, $a$=110 nm, $b$=110 nm, and the lattice constant of the array is 660 nm. The $Q$ factor increases rapidly as the array size is enlarged. It has a value larger than $10^3$ with an array size of 11×11. The high $Q$ factor would support the effective interaction between the resonance mode and silicon surface. We attribute the high $Q$ factor to the appearance of the quasi-BIC mode.[35],[38] A periodic array of nanocubes without air-slots can achieve the symmetry-protected BIC.[39] This mode can stably exist above the lightline with infinite $Q$ factor, as verified by the band structure calculation shown in the Supporting Information. Once the broken symmetry is introduced, the mode turns into a leaky mode with high $Q$ factor.[35][40] Here, the existence of the air-slot and substrate can be regarded as the broken symmetry to the symmetry-protected BIC, resulting in a quasi-BIC mode with high $Q$ factor in the slotted nanocube array.

The mode distribution of the resonance mode from an 11×11 array is displayed in Figure 2b, and the inset shows a zoomed image of the central slotted nanocube. A circulating electric field is still observed within each nanocube. Interestingly, the innermost nanocube shows the largest field enhancement while the outermost nanocubes show smaller field enhancements, which will be explained in the next section. Note the central nanocube has a significantly boosted electric field at the boundary of the air-slot. By comparing with an array of nanocubes without air-slot, the electric field at the air-slot boundary of the slotted nanocube array is about 8 times stronger than that inside the nanocube array without air-slot. The arrows in the inset display the electric field directions, indicating the electric fields across the air-slot are mainly the $x$-component. By increasing the array size, not only the electric field at the air-slot boundaries can be enhanced, more air-slot boundaries are generated for providing surface $\chi^{(2)}$ nonlinearity.

To further describe the surface SHG process of the silicon slotted nanocubes, the nonlinear polarization (i.e. $P_{\text{surface}}$) is expanded as three non-zero contributions:[27]

$$P^{(2)}_{\perp\perp\perp} = \chi^{(2)}_{\perp\perp\perp}\left[E^2_\perp\right]\hat{e}_\perp, \tag{1a}$$

$$P^{(2)}_{\perp\|\|} = \chi^{(2)}_{\perp\|\|}\left[E^2_\|\right]\hat{e}_\perp, \tag{1b}$$

$$P^{(2)}_{\|\|\perp} = \chi^{(2)}_{\|\|\perp}\left[E_\perp E_\|\right]\hat{e}_\|, \tag{1c}$$

where $\chi^{(2)}$ is the second-order nonlinear susceptibility, $\|$ and $\perp$ represent the directions parallel and perpendicular to the surface. According to Figure 2b, the electric field of the resonance mode around the air-slot is along the $x$ direction. Consequently, the enhancement in the SHG at the air-slot is dominated by $E_x$.

We then calculate the surface SHG from the periodic array of slotted nanocubes ($l$=440 nm, $a$=110 nm, $b$=110 nm, and lattice constant of 660 nm), which is enhanced by introducing the air-slot as well as the quasi-BIC state. For comparison, SHG excited from a periodic array of nanocubes without air-slots ($l$=440 nm and lattice constant of 660 nm) is also calculated. The SHG intensity is calculated as follows:[41]

$$I_{SHG} = \int_A \vec{S}_{SH} \cdot n\, da \tag{2}$$

where $\vec{S}_{SH}$ is the Poynting vector of the SHG field and $n$ is the unit vector normal to a surface $A$. SHGs from all surfaces of the nanocube are calculated and added together to obtain the overall SHG. Figure 2c displays the calculated SHG spectra from the arrays of nanocubes with and without air-slots. Beneficial from the extra air-slots, the SHG is enhanced by nearly 200 times from the periodic array of silicon nanocubes.

Like the perturbation of the symmetry-protected BIC,[39],[40] the dimension of the air-slot would determine the $Q$ factor of the BIC, which consequently modifies the enhanced electric field over the sidewall surface of the air-slot as well as the SHG process. In Figure 2d, we plot the enhancement factors of the electric field and the SHG as functions of the air-slot width $a$ with respect to the values obtained with $a$=0 nm, i.e., nanocubes without air-slots. Here, the considered electric field is defined as the average value over the sidewall surface of the air-slot (the red area in the inset). As $a$ decreases, the enhancement factor of the electric field approaches 300, and the SHG enhancement factor is increased to $10^6$ gradually. The reduction of $a$ leads to a situation that the radiation loss gradually decays for a symmetry-protected BIC mode. In consequence, there produces strong field enhancement thus corresponding to a strengthened SHG intensity.[19],[42] At $a$=0 nm, i.e. no air-slot, the SHG only originates from the weak surface effect at the nanocube boundary and hence has low efficiency. Moreover, we study the dependences of the $Q$ factors and SHG enhancements on other air-slot geometries, i.e., the air-slot depth and the shift of air-slot from the nanocube center, as discussed in the Supporting Information. With the decrease in the air-slot depth, both $Q$ factor and SHG enhancement are improved, which is similar to that with the varied air-slot width. It could be ascribed to that the variations of the air-slot depth or the air-slot width could both be considered as the changes in the perturbation of broken-symmetry, which should have equivalent effect to the supported quasi-BIC.[40] As for the fixed air-slot, the $Q$ factor and SHG enhancement can be improved by shifting the air-slot from the nanocube center gradually. Only for large air-slot shifts, the improvement in the $Q$ factor and SHG enhancement is significant. The simulation results confirm the crucial role of the electric field enhancement on the Si surfaces in the improvement of the SHG enhancement. These phenomena emphasize that the air-slot geometries are important in the proposed strategy.

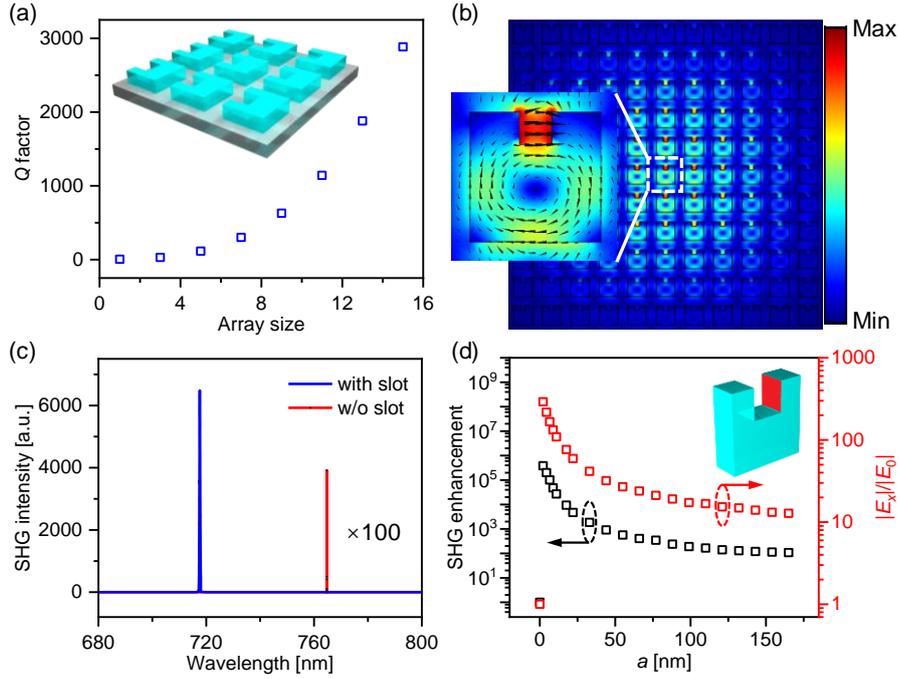

**Figure 2.** (a) Dependence of *Q* factors on the array size for the arrayed slotted nanocubes. Inset: schematic of a slotted nanocube array with a size of 3×3. (b) Electric field distribution in the slotted nanocubes with an array of 11×11. Inset: Electric field vector plot of the central slotted nanocube. (c) Comparison of the calculated SHG from the arrays of nanocubes with and without air-slots. (d) Enhancement factors of the electric field over the sidewall surface of the air-slot and the overall SHG as functions of the slot width *a*.

## 3. Experimental Results and Discussions

To experimentally verify our proposal, arrays of nanocubes with varied air-slot widths are fabricated on a silicon slab with the techniques of electron beam lithography and inductively coupled plasma etching (for details, see Experimental Section). The array of the slotted nanocubes is 20×20 with a lattice constant of 740 nm. The nanocube has a length of 480 nm. The air-slot widths are changed while the depths are fixed at 123 nm. Scanning electron microscope (SEM) images of a fabricated device are displayed in **Figure 3**a, showing decent surface morphology. We first characterize their linear optical responses by measuring their reflection spectra in a home-built microscope (see Experimental Section for details), as shown in Figure 3b. For the devices with different air-slot widths, all of them show prominent reflection peaks. The resonance wavelengths locate at 1551.8 nm, 1544.8 nm, 1539.3 nm, 1533.4 nm, and 1531.5 nm, respectively, for the device with the air-slot widths of 116 nm, 134 nm, 153 nm, 184 nm, 194 nm. The *Q* factors of these resonances have values of 651, 489, 440, 365, and 300. The *Q* factors reduce with the increase of air-slot widths, accompanied by a blue shift in the resonance wavelength. Note that the measured *Q* here is a combination of the radiative ($Q_r$) and nonradiative ($Q_{nr}$) contributions.[32] As a structure supporting the symmetry-protected BIC, $Q_r$ can be adjusted by the broken-symmetry perturbation. However. the existence of the nonradiative losses, including non-ideal factors caused by the fabrication process and the finite array, leads to the attenuation in the measured *Q* factor (for details, see Supporting Information). Considering that these two contributions are inevitable in the experiments, the measured results manifest themselves as lower *Q* factors compared with the calculated ones.

According to Figure 2b, the *x* component of the electric field can be enhanced due to the continuous condition of electric displacement across the air-slot boundary. It means that the *x*-polarized component can inspire the resonance mode effectively. To confirm this point, we measure the reflection intensity of the sample with *a*=116 nm by rotating the polarization of the on-resonance laser, as shown in Figure 3c. A linear polarization dependence along the *x*-direction with a ratio of 0.76 is obtained. Figure 3d displays the spatial distribution of the resonance mode examined under the far-field excitation and collection. The mode localizes around the center of the array. In an infinite array with the symmetry-protected BIC, the normal radiative decay of the mode is compensated by driving terms arising from the local field at the positions of the nanocubes.[43],[44] However, as a finite array, nanocubes around the boundary can only be partially compensated, resulting in an extra leakage. Thereby the center area of the structure has a stronger ability of field confinement. Notably, this extra leakage is inevitable for a practical device and finally leads to the decline in the *Q* factor.

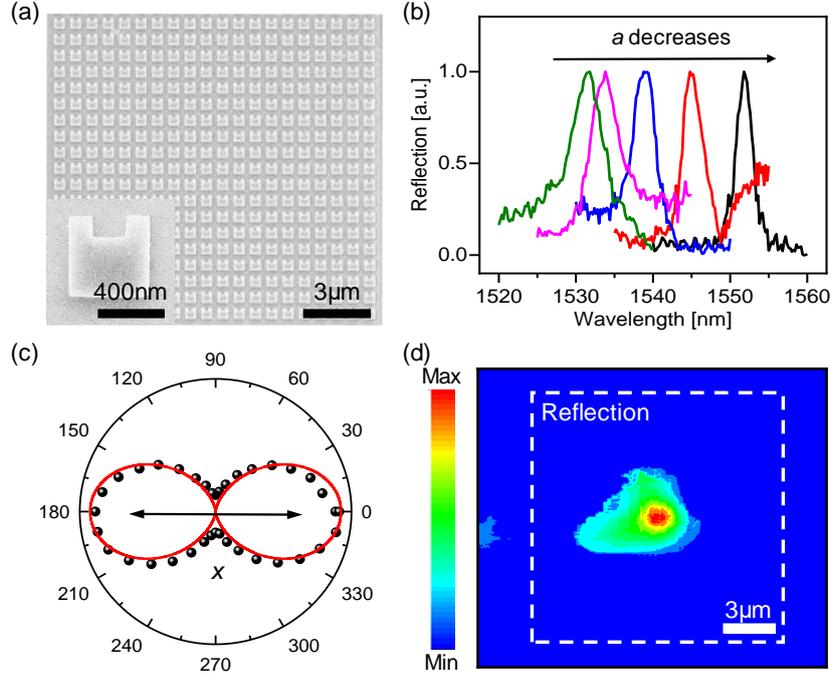

**Figure 3.** (a) SEM images of the fabricated slotted nanocube array. (b) Measured reflection spectra of the slotted nanocube arrays with different air-slot widths. (c) Normalized polarization dependence of the reflection intensity from a slotted nanocube array with the excitation of an on-resonance laser. (d) Spatial mappings of the mode distribution with the excitation of an on-resonance laser. The white dashed frame implies the boundary of the sample.

SHG processes from the fabricated devices are then characterized using a reflection system (see Experimental Section for details). **Figure 4**a displays the measured SHG spectra from two arrays of the nanocubes with and without air-slots. There is no detectable SHG from the structure without air-slots because the SHG signal is so weak that it is out of the measurement sensitivity of our system. On the contrary, a significant SHG signal can be observed in the structure with slotted nanocubes ($a$=116 nm). SHG is expected to show a quadratic dependence on the pump power. In Figure 4b, we display a log-log plot of the SHG versus the pump power at the resonance wavelength. The fitted slope is 1.82±0.03, indicating a typical quadratic relation between the pump and signal in the SHG process. The polarization dependence of the SHG is shown in Figure 4c. The SHG intensities reach the maximal values for $x$-polarized excitation, which has the same polarization dependence as that in the reflection spectrum shown in Figure 3c. Note the extinction ratio of the SHG polarization dependence (i.e. 0.98) is much higher than that of the resonance mode. It is because an $x$-polarized pump light can inspire the resonance mode effectively and the SHG process has a quadratic function of the pump light. The spatial mapping of the SHG signal with the on-resonance pump is also carried out, as shown in Figure 4d. A similar spatial distribution to that of the resonance mode is observed because the SHG is excited by the resonance mode. Note, the size of the SHG distribution is smaller than that of the resonance mode considering the quadratic function in the SHG process between the pump laser and the SHG signal.

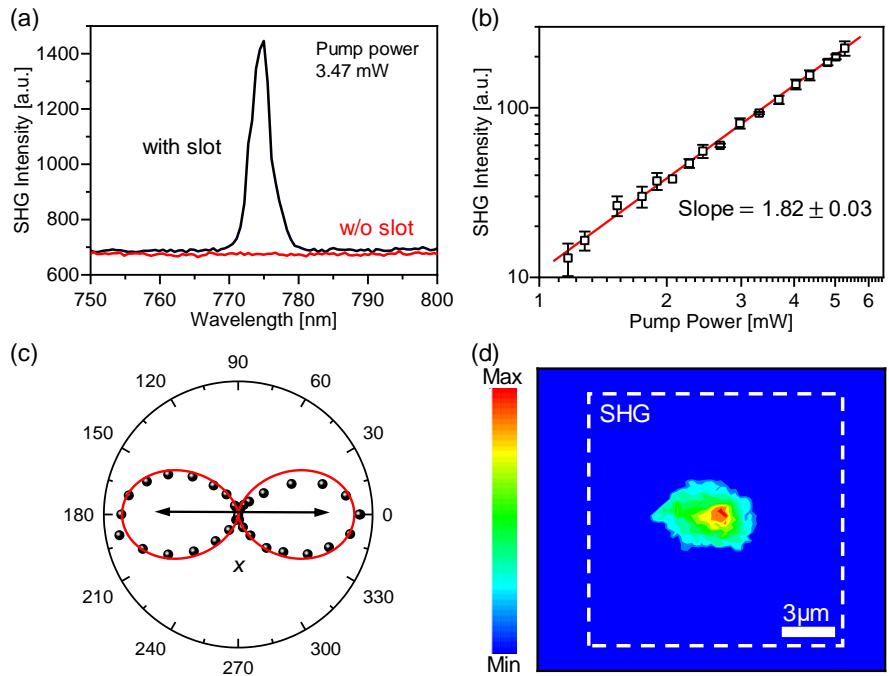

**Figure 4.** (a) Experimentally measured SHG spectra from the arrays of nanocubes with and without air-slots. (b) Log-log plot of SHG power-dependence on the pump powers (black dots) and their linear fitting (red curve). (c) Normalized polar plots of the SHG signals (black dots) and their fitting (red curve). (d) Spatial mapping of the SHG pumped with the on-resonance laser.

The link between the resonance-enhanced SHG and the resonance mode of the slotted nanocube array is visualized by the pump-wavelength dependence of the SHG signals. Here, a picosecond laser with the tunable wavelength range of 1530-1560 nm is employed. It is focused onto a sample with $a$=153 nm at the resonance wavelength of 1539.3 nm. **Figure 5**a displays the SHG spectra obtained with the pump lasers at different wavelengths. When the pump wavelength is scanned across the resonance wavelength, SHG reaches the maximum at the resonance wavelength and dramatically decreases under the off-resonant conditions. In Figure 5b, we plot the dependence of the SHG intensity on the pump wavelength. The experiment result is well fitted by the quadratic Fano function because the SHG is a resonance-enhanced nonlinear frequency conversion process. The near field distribution of the structure is calculated at different wavelengths, as shown in Figure S4 of the Supporting Information. At the resonance wavelength, there is a significant improvement in the electric field enhancement at the surface of the slotted nanocube. The intensity of the electric field decays rapidly once the wavelength is shifted away from the resonance wavelength. Such surface overlapping of the resonance modes supported by the slotted nanocubes can lead to the SHG enhancement compared with the off-resonance case. It is confirmed that the SHG enhancement originates from the field concentration at the quasi-BIC mode. As discussed above, the air-slot-assisted SHG enhancement can be governed by varying the air-slot width ($l$=480 nm, $b$=123 nm, and the lattice constant is 740 nm). To verify it, we measure SHG intensities from the devices with varied air-slot widths, as shown in Figure 5c. With the decrease in $a$, the SHG intensities gradually increase, which is in agreement with the dependence of the calculated results shown in Figure 2d. Finally, we estimate the SHG conversion efficiency in a structure ($l$=515 nm, $a$=117 nm, $b$=101 nm, and the lattice constant of 632 nm), defined as $\eta=P_{SHG}/P^2_{pump}$. With a pump power of 3.74 mW, the SHG with a power of 2.53 nW was measured. The normalized conversion efficiency is $1.8\times10^{-4}$ W$^{-1}$ (~$10^{-14}$ W$^{-1}$ in bulk silicon[45]), which is larger than that obtained from the photonic crystal nanocavity based on Si.[26] Compared with silicon nanoparticles,[25] our structure can achieve a comparable SHG efficiency ($6.8\times10^{-7}$) at a much lower pump peak intensity (~0.5 GW/cm$^2$).[46] It should be pointed out that the SHG efficiency can be affected by the pulse duration (for details, see Supporting Information). With the same pulse energy, a shorter pulse duration would result in a higher SHG efficiency. Enhancing surface SHGs with nanoslots or nanogaps have also been demonstrated in metal plasmonic nanostructures,[47]-[55] such as bow-ties,[47] nanogrooves,[50] nanocube arrays.[53] Compared with them, the SHG conversion efficiency achieved in our proposed array of Si slotted nanocubes can be several orders of magnitude higher (see **Table S1** in Supporting Information). In addition, given the plasmonic structures suffer from high losses and Joule heat, the Si-based structure displays the advantages of low loss and CMOS-compatible fabrication process.

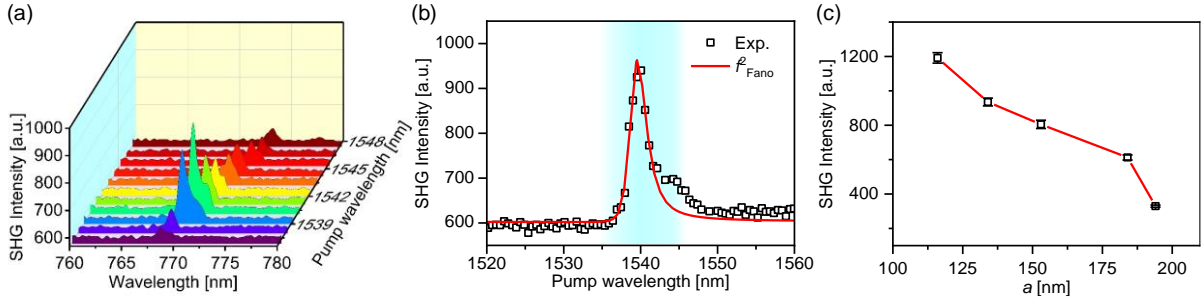

**Figure 5.** (a) Evolution of the SHG spectra at different pump wavelengths. (b) Wavelength dependence of SHG intensities when the pump wavelength is tuned across the resonance wavelength, where $f_{Fano}$ is the Fano function used to fit the resonance wavelength (see details in Experimental Section). The blue area represents the resonance-enhanced SHG. (c) Dependence of the measured SHG intensities on the varied air-slot widths.

## 4. Conclusion

In conclusion, an array of silicon slotted nanocubes is designed and fabricated to boost the second-order nonlinearity from the centrosymmetric silicon. The air-slot in the silicon nanocube could provide enlarged surface area to utilize the surface second-order nonlinearity. On the other hand, the air-slot promises a significantly enhanced optical field due to the continuous condition of the normal component of the electric displacement. Further, the slotted nanocubes are arranged into an array to form the quasi-BIC with high $Q$ factor, which ensures the effective light-matter interaction in excitation the surface second-order nonlinearity of silicon. Consequently, SHG from the slotted nanocube array is improved by more than two orders of magnitude compared with that from the array of silicon nanocubes without air-slots. The experimentally measured SHG efficiency of the silicon slotted nanocube array is high as $1.8 \times 10^{-4}$ $W^{-1}$, which could be further engineered by modifying the air-slot geometries, such as the width and depth of the air-slot, and the shift of the air-slot from the nanocube center. Our results could provide a new strategy to develop second-order nonlinear optical effects and devices with high efficiency from centrosymmetric materials.

## 5. Experimental Section

*Numerical Simulations*: Numerical calculations were carried out using the finite-element method (COMSOL Multiphysics). The silicon nanocube ($n$=3.48) placed on a sapphire substrate ($n$=1.75) was illuminated by the normal incident plane wave with a polarization along the *x*-axis. Perfectly matching layers were used at the input and output ports. The quality factors were evaluated by the eigenmode solver. For a periodic array of the silicon nanocubes, periodic boundary conditions were applied along the *x*/*y* directions.[56] Second harmonic generation from the slotted nanocubes is modeled through two electromagnetic simulation steps in the frequency domain. First, the linear scatterings at the pump wavelength are simulated, and the induced nonlinear polarizations are derived. Second, these nonlinear polarizations are set as the sources for the SHG fields at the doubled frequency. In all simulations, the nanocube thickness $t$ is fixed at 230 nm.

*Sample Fabrication and Characterization*: The sample used for the device fabrication is a silicon-on-sapphire wafer with a 230 nm thick silicon layer. The wafer was diced in 1 cm×1 cm pieces, which were cleaned in *N*-methyl-2-pyrrolidone, isopropyl alcohol, and Deionized water for 10 minutes. After that, AR-P 6200.13 was spun on the sample at 4000 rpm to have a thickness of about 400 nm. The resist was then baked on a hot plate at 180 °C for 2 minutes. Electron beam lithography was performed on the samples to transfer the pattern on the resist. Subsequently, the pattern was transferred into the silicon using inductive coupled plasma with $SF_6$ and $C_4F_8$ as etchants. Finally, the residual resist was chemically removed with *N*-methyl-2-pyrrolidone and piranha solution. The morphology of samples was assessed by a scanning electron microscope (SEM, FEI Verios G4).

*Optical Characterizations*: The samples were characterized using a home-built setup, which has both reflection and transmission configurations. For linear optical responses, samples were illuminated by the light from a broadband supercontinuum laser (YSL photonics, SC-Pro). The spectra of reflected light were collected by means of an objective (50×, NA=0.42), which was subsequently coupled to an infrared spectrometer (Andor tech., DU490A-1.7). Moreover, to evaluate the *Q* factors of the resonance modes accurately, samples were illuminated by a narrowband tunable laser (Yenista Tunics T100S-HP), and reflected signals were measured using a telecom-band photodiode. The measured data were fitted by the Fano function $f_{Fano}=((q+(\omega-\omega_0)/\gamma)^2+\gamma^2)/(1+((\omega-\omega_0)/\gamma)^2)$ to obtain the *Q* factor.[57] Here $q$ is the asymmetry factor, $\omega_0$ is the resonance frequency, and $\gamma$ represents the damping rate. The SHG is collected by a 20×/0.4 NA microscope objective when pumped by a fiber-based pulsed laser with a pulse width of 8.8 ps and a repetition rate of 18.5 MHz. The signals were then coupled to a visible spectrometer (Princeton Instruments, SP 2558 & 100BRX). The powers of the transmitted SHG are measured by a visible photomultiplier tube. During the measurement, a polarizer and a half-wave plate were utilized to control the linear polarization of the pump laser illuminated on the sample. The spatial mappings are measured using a setup with a reflection configuration. The sample was mounted on a two-dimensional (2D) piezo-actuated stage. To characterize the resonance mode, a narrowband tunable laser (Yenista Tunics T100S-HP) with an on-resonance wavelength is focused onto the sample using a 50×/0.42 NA microscope objective. The reflected powers are monitored using a telecom-band photodiode. As for the far-field radiations of the SHG, they were monitored using a visible spectrometer (Princeton Instruments, SP 2558 & 100BRX) when a pico-second laser was utilized to generate the fundamental beam with the on-resonance condition (pulse duration 8.8 ps and repetition rate 18.8 MHz). Homemade LabVIEW programs were designed to synchronously control the 2D piezo-actuated stage and a telecom-band photodiode/visible spectrometer for measuring the spatial mappings of the structure.

**Supporting Information**

Supporting Information is available from the Wiley Online Library or from the author

**Acknowledgments**

The authors acknowledge support from the National Key Research and Development Project (Grant No. 2018YFB2200500), the National Natural Science Foundation of China (Grant No. 62025402, 62090033, 91964202, 92064003, 61874081, 61851406, 62004149, 62004145, 91950119, and 61775183), and Major Scientific Research Project of Zhejiang Lab (No. 2021MD0AC01).

[52]	J. Deng, Y. Tang, S. Chen, K. Li, A. V. Zayats, G. Li, *Nano Lett.* **2020**, *20*, 5421.

[53]	Y. Zeng, H. L. Qian, M. J. Rozin, Z. W. Liu, A. R. Tao, *Adv. Funct. Mater.* **2018**, *28*, 1803019.

[54]	X. Wu, W. Jiang, X. Wang, L. Zhao, J. Shi, S. Zhang, X. Sui, Z. Chen, W. Du, J. Shi, Q. Liu, Q. Zhang, Y. Zhang, X. Liu, *ACS Nano* **2021**, *15*, 1291.

[55]	G. -C. Li, D. Lei, M. Qiu, W. Jin, S. Lan, A. V. Zayats, *Nature Commun.* **2021**, *12*, 4326.

[56]	C. Fang, Y. Liu, G. Han, Y. Shao, J. Zhang, H. Yue, *Opt. Exp.* **2018**, *26*, 27683.

[57]	I. Avrutsky, R. Gibson, J. Sears, G. Khitrova, H. M. Gibbs, J. Hendrickson, *Phys. Rev. B* **2013**, *87*, 125118.

An array of slotted nanocubes is proposed to achieve strong SHG effects in Si. BIC-induced electric field enhancement and slot-induced surface effect play a crucial role in the SHG. The high conversion efficiency of $1.8\times10^{-4}$ $W^{-1}$ can be experimentally obtained. This strategy can boost efficient SHG even in centrosymmetric materials

Cizhe Fang, Qiyu Yang, Qingchen Yuan, Linpeng Gu, Xuetao Gan,* Yao Shao, Yan Liu,* Genquan Han, Yue Hao

**Efficient Second-Harmonic Generation from Silicon Slotted Nanocubes with Bound States in the Continuum**

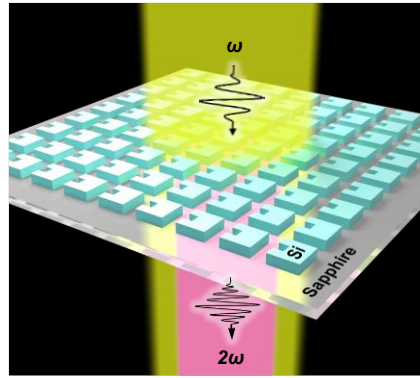

# Supporting Information

**Efficient Second Harmonic Generation from Silicon Slotted Nanocubes with Bound States in the Continuum**

*Cizhe Fang, Qiyu Yang, Qingchen Yuan, Linpeng Gu, Xuetao Gan,\* Yao Shao, Yan Liu,\* Genquan Han, and Yue Hao*

C. Fang, Q. Yang, Prof. Y. Liu, Prof. G. Han, Prof. Y. Hao
Wide Bandgap Semiconductor Technology Disciplines State Key Laboratory, School of Microelectronics, Xidian University, Xi'an, 710071, China
E-mail: xdliuyan@xidian.edu.cn

Q. Yuan, L. Gu, Prof. X. Gan
Key Laboratory of Light Field Manipulation and Information Acquisition, Ministry of Industry and Information Technology, and Shaanxi Key Laboratory of Optical Information Technology, School of Physical Science and Technology, Northwestern Polytechnical University, 710129 Xi'an, China
E-mail: xuetaogan@nwpu.edu.cn

Y. Shao
Shanghai Energy Internet Research Institute of State Grid, 251 Libing Road, Pudong New Area, Shanghai, 201210, China

Prof. Y. Liu
The Research Center for Intelligent Chips and Devices Zhejiang Lab, Hangzhou, 311121, China

## I. Explaination of BICs in the array of nanocubes.

We consider a periodic array of nanocubes on a sapphire substrate (**Figure 1**a in the maintext), where a lattice (periodicity, $p$, is set as 660 nm) of nanocubes (side length $l$=440 nm) is patterned in a silicon slab (thickness $t$=230 nm). The band structure calculated by COMSOL Multiphysics is shown in **Figure S1**(a). The mode can exist stably above the lightline, implying that the structure can support a BIC mode at the $\Gamma$ point of the first Brillouin zone. The inset shows that the electric field manifests itself as the circular current. The quality ($Q$) factor of the mode is modeled as a function of $k_x$ (the broken symmetry) and $l$ (the geometric parameter), respectively. As shown in Figure S1(b), the mode has a nearly infinite $Q$ factor (more than $10^9$) at $k_x$=0. The $Q$ factor drops sharply once $k_x$ is away from the $\Gamma$ point as a result of symmetry mismatch. It indicates that the mode is sensitive to the broken-symmetry perturbation. By contrast, this mode is insensitive to geometrical change since the $Q$ factor preserves ultrahigh value (i.e. more than $10^8$) with the change in $l$. Above all, the proposed periodic array of nanocubes can support a symmetry-protected BIC.

The distribution of the $Q$ factor in momentum space for the nanocube array without a substrate is shown in Figure S1(c), featuring the BIC where $Q$ diverges to infinity at the $\Gamma$ point. Notably, the $Q$ factor at the $\Gamma$ point for a structure without the substrate is more than $10^{10}$. It means that the BIC can get suppressed due to the existence of substrate. Actually, the mode shown in Figure S1(a) can be considered as a quasi-BIC mode. The topological nature of the BICs can be understood from the evolution of topological charges in Figure S1(d), where the BIC appears as vortices of the polarization vector and is characterized by an integer topological charge of +1.[S1]-[S3] Robust BICs are only possible with the existence of vorticity in the polarization field. The topological charge represents times of the polarization vector winds around the BIC. It is pinned at the center of the Brillouin zone owing to symmetry.

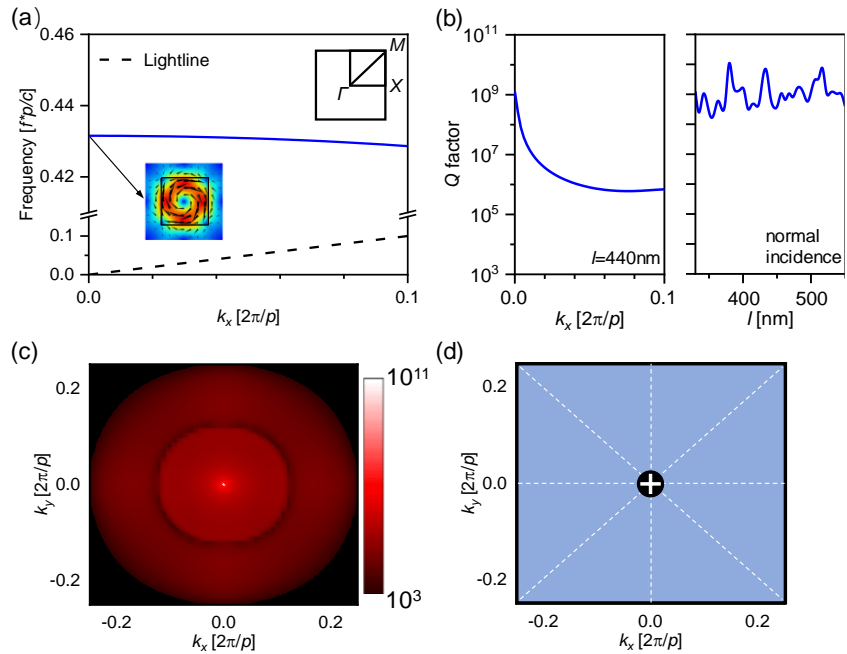

**Figure S1.** (a) Band diagram of the silicon nanocube in a square lattice. Inset: the mode distribution of the nanocube in the array. (b) Corresponding $Q$ factor as functions of $k_x$ (left

panel) and *l* (right panel), respectively. (c) Evolution of the calculated *Q* factors in the momentum space for the nanocube array without a substrate. (d) Corresponding topological configuration.

## II. Numerical investigations of the dependences of the slot geometries

To clearly show the effect of the air-slot geometries on the quasi-BIC mode and SHG enhancement, we discuss the simulated results below.

1) Effect of air-slot depth *b* on quasi-BIC and SHG enhancement

Extensive numerical simulations are carried out by varying *b*, where *l*, *a*, and *p* are fixed at 440 nm, 110 nm, and 660 nm, respectively. The simulation results are shown in **Figure S2**(a). From the figure, we can conclude:

  i.    The *Q* factor increases with the decrease in *b*;

  ii.   The SHG enhancement increases with the decrease in *b*.

The effect of *b* on the *Q* factor and SHG enhancement is similar to that of the air-slot width *a* as discussed in the maintext. It is because, for the variation of the broken-symmetry perturbation (i.e. the air-slot area), the effect of the air-slot depth is equivalent to the air-slot width. The reduction of *b* can be simply understood as the decline in broken-symmetry perturbation. With the decrease in *b*, the *Q* factor of the mode becomes larger, leading to the enhancement in the electric field intensity at the sidewall surface of the air-slot and thus the SHG intensity (see Figure S2(b)). Moreover, the electric field at the other sidewall surfaces can also be enhanced, as shown in the inset of Figure S2(b). To improve the SHG efficiency, the air-slot depth *b* should be as small as possible.

2) Effect of the shift of the air-slot from the nanocube center *g* on quasi-BIC and SHG enhancement

In the simulation, *l*, *a*, *b*, and *p* are fixed at 440 nm, 110 nm, 110 nm, and 660 nm, respectively, and the air-slot is shifted from the nanocube center gradually with a distance of *g*. The simulation results are shown in Figure S2(c).

i.  The *Q* factor increases along with the enlarged *g*;

ii. The SHG enhancement increases along with the enlarged *g*.

According to Figure S2(d), with the increase in the air-slot shift *g*, the intensity of the electric field is gradually enhanced, especially at the sidewall surfaces of the air-slot. We attribute this phenomenon to the mode mixing between the transverse and the longitudinal dipole modes. The decay of the transverse mode (i.e. the electric dipole mode) is governed by both radiative and nonradiative processes while the longitudinal mode (i.e. the magnetic dipole mode) only decays due to nonradiative losses. As *g* increases, the in-plane electric field mode gradually couples with the longitudinal magnetic dipole (see the top inset of Figure S2(c)). It leads to the decay in the mode loss. It is worth mentioning that only for a large value of *g*, the *Q* factor can be significantly improved as well as the SHG enhancement. It may be difficult for device fabrication. Above all, the air-slot shift *g* can be a good choice to improve the *Q* factor and SHG enhancement when the slot geometries are fixed.

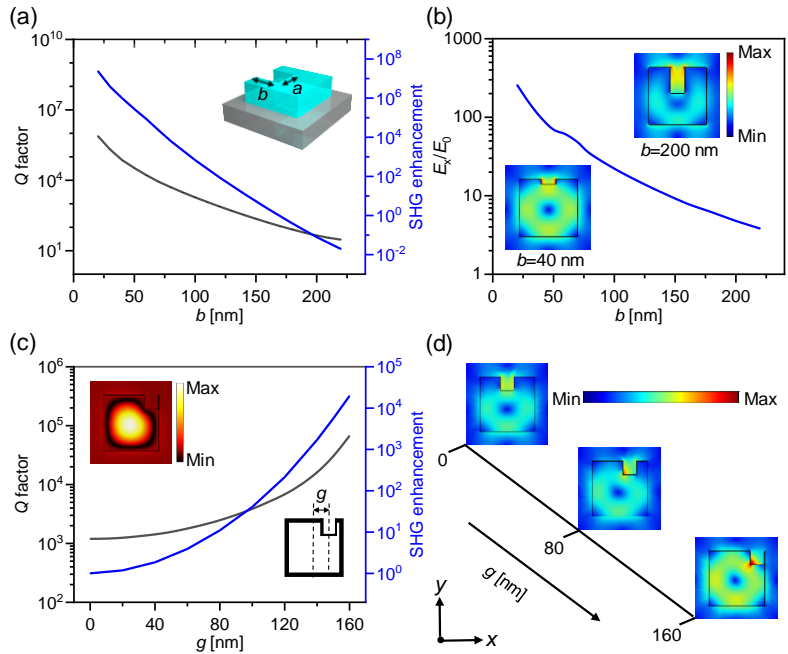

**Figure S2.** (a) *Q* factor and the SHG enhancement as a function of the air-slot depth *b*. The overall SHG from the structure with *b*=0 nm is used as a reference. (b) Electric field enhancement over the sidewall surface of the air-slot as a function of the air-slot depth *b*. The inset shows the electric field intensity distributions at *b*=40 nm and 200 nm. (c) *Q* factor and the SHG enhancement as a function of the air-slot shift *g*. The overall SHG from the structure

with $g$=0 nm is used as a reference. The top inset shows the magnetic field intensity distribution at $g$=160 nm. The bottom inset shows the air-slot shift from the center, denoted by $g$. (d) Electric field intensity distributions at $g$=0 nm, 80 nm, and 160 nm, showing the gradually strengthened electric field on the side-walls.

**III. Analysis of the *Q* factor in the slotted nanocube array**

In the experiments, the total *Q* factor can be characterized via $1/Q=1/Q_r+1/Q_{nr}$. As a structure supporting the symmetry-protected BIC, $Q_r$ can be adjusted by the broken-symmetry perturbation. However, the existence of the nonradiative factor leads to the difference between the calculated value and measured value (see **Figure S3**(a)). The main nonradiative contributions to the low-*Q* resonance in our structures are the non-ideal factors, caused by the device fabrication (including the surface roughness,[S4] cleanliness, and structure disorder), and the finite array. As shown in Figure. S3(b) and (c), we compared the SEM images and the measured reflection spectra of the samples before and after the process optimization. It is clear that for a structure after the process optimization, the surface is cleaner and the edges of the blocks are smoother than those in the structure without the process optimization. Importantly, the *Q* factor can be significantly improved. No impurities and structure disorders can minimize the influences of non-ideal factors caused by the device fabrication.

Another nonradiative contribution results from the finite array. It is also inevitable in the experiments. To interpret the relationship between the *Q* factor and the array size, the electric field intensity distributions in different sizes of the nanocube array are simulated with identical parameters as the sample measured in the experiments (see Figure S3(d)). As we mentioned in the main text, the center area of the structure has a stronger ability of field confinement due to the compensation effect assisted by the local field at the positions of the nanocubes. With the increase in the array size, the proportion of the unit cell at the edge can be reduced and the center area can produce larger field enhancement, corresponding to less leakage (i.e. larger *Q* factor), as shown in Figure S3(e). Since the array size can not increase infinitely, a reasonable scheme can be the best choice for a finite array to improve the *Q* factor.

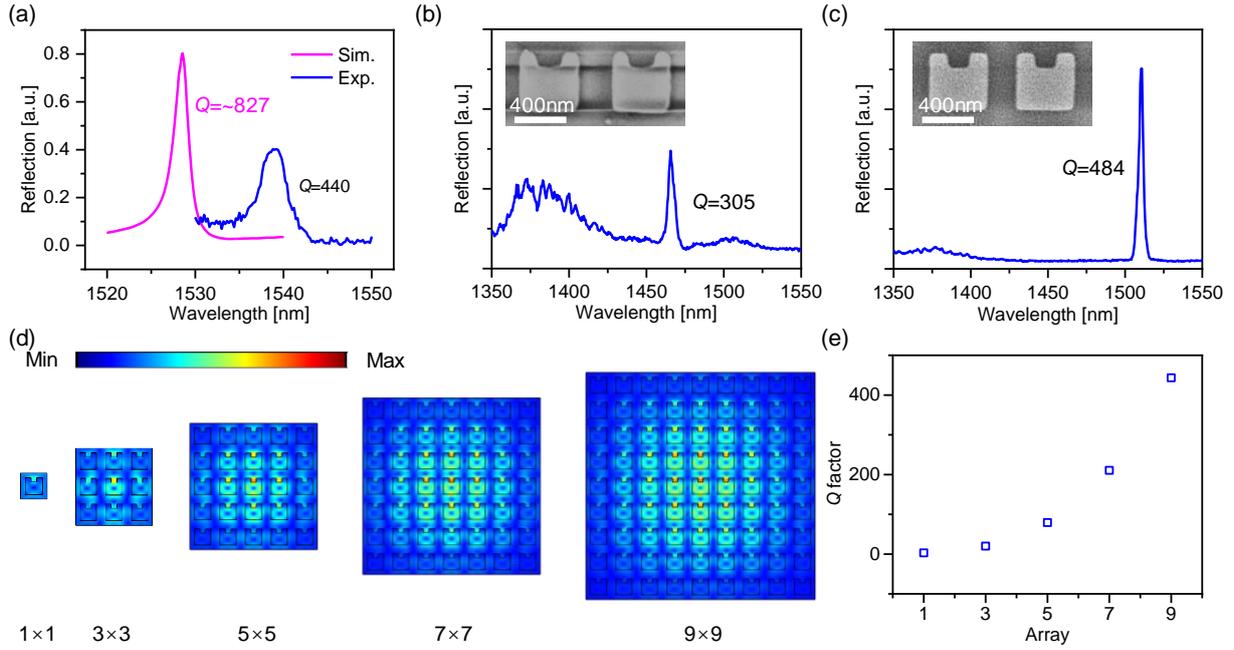

**Figure S3.** (a) Numerically calculated and the experimentally measured reflection spectra for the structure with geometric parameters of $l$=480 nm, $a$=153 nm, $b$=123 nm, and the lattice constant is 740 nm. (b,c) A comparison of the SEM images and measured reflection spectra of the samples before (b) and after (c) the process optimization. (d) Simulated electric field distributions obtained at the vertical mid-plane of the nanocubes with different array sizes. (e) Calculated $Q$ factors from slotted nanocube arrays with different sizes.

## IV. Calculation of the near-field distributions of the resonance mode in the slotted nanocube

For qualitative research on the effect of the field enhancement on SHG, we performed a numerical simulation of the near-field distribution of the resonance mode in the slotted nanocubes. The structure has parameters of $l$=480 nm, $a$=153 nm, $b$=123 nm, and $p$=740 nm. The calculated reflection spectrum is shown in **Figure S4**(a). The field distributions of one unit in the array at different wavelengths (i.e. A, B, and C) are also simulated and shown in Figure S4(b). At the peak wavelength (point C), the electric field enhancement can be significantly improved. In particular, the maximum field concentration occurs at the sidewall of the air-slot surface. There is a significant attenuation in the electric field once the wavelength is away from the resonance wavelength, consistent with the dependence of the SHG intensity on the pump wavelength shown in Figure 5(b). Given the electric field magnitude of the SHG is proportional to the square of the fundamental optical field, it is evident that the enhancement of the SHG originates from the field enhancement at the quasi-BIC mode.

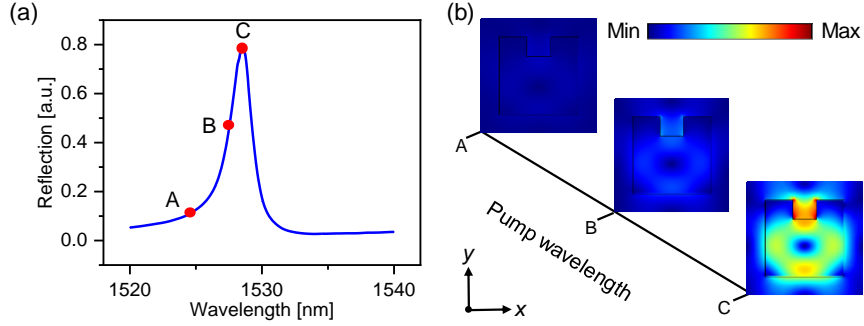

**Figure S4.** (a) Calculated reflection spectrum for the structure employed in the experiment shown in the maintext. (b) Near-field distributions of the electric field at the wavelengths indicated by points A, B, and C in (a).

## V. Qualitative analysis on the effect of the pulse duration on the SHG efficiency

Besides the specifically designed silicon nanostructure, the SHG efficiency of the silicon slotted nanocube array can also be affected by the parameters of the employed pump pulses. We present a theoretical analysis of the dependence of the SHG efficiency on the pulse duration. The pulsed laser used in this work has Gaussian pulses with a pulse duration $\tau$=8.8 ps and a repetition rate $Rep$=18.5 MHz. The time-function of the pulse power could be written as

$$P(t) = P_0 \exp(-2t^2/\tau^2) \tag{1}$$

where $P_0$ is the peak power of a single pulse. For simplicity, we assume the energy in each pulse is a constant and set it as $E$.

$$E = \int P(t)\,dt \tag{2}$$

For the averaged power $P_{ave}$, the corresponding relationship between the peak power $P_0$ and pulse duration $\tau$ can be

$$Rep \times \int P(t)\,dt = P_{ave} \times 1\text{s} \tag{3}$$

During the second order nonlinear process, the SHG power $P_{SHG}$ is proportional to the squared pump power $P_{pump}$ (i.e., $P_{SHG} \propto P^2_{pump}$). For the pulsed laser, it should be proportional to $P^2(t)$.

For the employed pulsed laser, in a time duration of 1 s, the energy of SHG is proportional to $Rep \times \int P^2(t)dt$. Combining with Eqs. (1)-(2), the SHG energy in 1 s generated by the pulsed laser can be characterized via

$$P_{SHG} \propto Rep \cdot \frac{E^2}{\sqrt{\pi}} \cdot \frac{1}{\tau}$$

In the case of the constant pulse energy, the SHG efficiency is inversely proportional to the duration of the pump pulses. The result indicates that the SHG efficiency can be further improved by engineering the parameters of the employed pump pulses.

**VI. Comparison of the SHG efficiencies with other plasmonic nanostructures**

There are many reported works about improving SHG in metals using plasmonic nanostructures with slots or asymmetries. It is better to do a comparison between them with our work, which studied the boosted surface SHG in silicon with an array of slotted nanocubes. SHG is forbidden in the bulk of metals due to the centrosymmetry. The surface SHG is possible from metals, and the employment of plasmonic nanostructures with nanogaps, nanoslots, nanoparticles could produce strong field confinements at the metal surfaces to achieve strong surface SHG effects. We compare the SHG conversion efficiency in these works with our work, as shown in **Table S1**. The metal plasmonic structures include the bowtie, nanogroove, trapezoidal antenna, nanocube, various nanoparticles, and hybrid structures. The SHG conversion efficiency in our proposed silicon array of slotted nanocubes is much larger than the corresponding values in these works. Given the plasmonic structures suffer from high losses and Joule heat, our silicon-based structure displays the advantages of low loss and preparation simplicity.

**Table S1.** Comparison of the SHG efficiencies in different plasmonic nanostructures with our results shown in the maintext.

| Year | Ref. | Material | SHG efficiency |
|---|---|---|---|
| 2012 | 49 | Ag (bowtie aperture)/$Si_3N_4$ | $1.4 \times 10^{-8}$ (0.42 GW/cm$^2$) |
| 2012 | 50 | Ag (three-arm trapezoidal antenna) | $10^{-9}$ (1.9 GW/cm$^2$) |
| 2015 | 51 | Au (nanoparticles) | $1.8\times10^{-7}$ (35.4 GW/cm$^2$) |
| 2015 | 52 | Ag (nanogroove array) | $1.2\times10^{-8}$ (14.6 GW/cm$^2$) |
| 2019 | 53 | Au (nanoparticle array) | $1.5\times10^{-9}$ (3.2 KW/cm$^2$) |
| 2020 | 54 | Au meta-atoms/ITO | $7.8\times10^{-8}$ W$^{-1}$ |
| 2020 | 55 | Au nanocube/Au film | $5.36\times10^{-9}$ (15.6 GW/cm$^2$) |
| 2021 | 56 | Au NPs/Al$_2$O$_3$/Al | $3.84\times10^{-9}$ (7 GW/cm$^2$) |
| 2021 | 57 | CTAB-coated Au NPs/Au | $3.56\times10^{-7}$ W$^{-1}$ |
| 2022 | This work | Si (metasurface) | $1.8\times10^{-4}$ W$^{-1}$ / $6.8\times10^{-7}$ (0.5 GW/cm$^2$) |